\documentclass{Interspeech2024}

\usepackage{enumitem}
\usepackage{multirow}
\usepackage{moresize}
\usepackage{amsmath}
\usepackage{lipsum}

\interspeechcameraready

\title{AnoPatch: Towards Better Consistency in Machine Anomalous Sound Detection}

\name[affiliation={1}]{Anbai}{Jiang}
\name[affiliation={2}]{Bing}{Han}
\name[affiliation={3}]{Zhiqiang}{Lv}
\name[affiliation={3}]{Yufeng}{Deng}
\name[affiliation={1,*}]{Wei-Qiang}{Zhang}
\name[affiliation={2}]{Xie}{Chen}
\name[affiliation={2}]{Yanmin}{Qian}
\name[affiliation={1,3}]{Jia}{Liu}
\name[affiliation={1,*}]{Pingyi}{Fan}

\newcommand\blfootnote[1]{%
  \begingroup
  \renewcommand\thefootnote{}\footnote{#1}%
  \addtocounter{footnote}{-1}%
  \endgroup
}

\address{
  $^1$Tsinghua University, China
  $^2$Shanghai Jiao Tong University, China \\
  $^3$Huakong AI Plus, China}
\email{jab22@mails.tsinghua.edu.cn, fpy@tsinghua.edu.cn}


\begin{document}

\maketitle

\begin{abstract}
  Large pre-trained models have demonstrated dominant performances in multiple areas, where the consistency between pre-training and fine-tuning is the key to success. However, few works reported satisfactory results of pre-trained models for the machine anomalous sound detection (ASD) task. This may be caused by the inconsistency of the pre-trained model and the inductive bias of machine audio, resulting in inconsistency in data and architecture. Thus, we propose AnoPatch which utilizes a ViT backbone pre-trained on AudioSet and fine-tunes it on machine audio. It is believed that machine audio is more related to audio datasets than speech datasets, and modeling it from patch level suits the sparsity of machine audio. As a result, AnoPatch showcases state-of-the-art (SOTA) performances on the DCASE 2020 ASD dataset and the DCASE 2023 ASD dataset. We also compare multiple pre-trained models and empirically demonstrate that better consistency yields considerable improvement.
\end{abstract}

\section{Introduction}
\label{sec:intro}

Machine anomalous sound detection (ASD) aims to predict whether the sound of a machine is normal or anomalous when only normal sounds are provided as prior. Recent years saw a growing need for robust ASD methods in the field of modern manufacturing, where machine audio under normal conditions can be utilized to detect unknown malfunctions, which has been demonstrated in some scenarios to significantly reduce risks and boost production efficiency.\blfootnote{* Corresponding authors}

Like any binary classification task, the machine ASD task has to balance between recall and precision. On one hand, to increase recall, the model is required to overfit with the normal distribution such that any deviation from the normal distribution yields a perceivable variation in the output. On the other hand, to increase precision, the model is required to grasp the universal features of normal samples and generalize well on unseen samples. However, these requirements become quite challenging for the ASD model, since the decision boundary can not be accurately determined when (i) only one class is presented for training and (ii) the normal distribution is sparse.

To implicitly derive the ASD boundary, training the model by deputy tasks is a common solution in previous literature\cite{wilkinghoff2021sub,liu2022anomalous,hou2023decoupling}. However, since the normal distribution is sparse and the training data is insufficient, training the model from scratch tends to yield non-scalable boundaries, especially when the machine has diverse working conditions. Thus, we opt to initialize the model by the parameters of a large pre-trained model that has obtained common knowledge of speech and audio by pre-training on extensive data, which enables the model to learn scalable boundaries when fine-tuned on machine audio.

Specifically, we focus on pre-trained models that are originally intended for audio classification and model the audio from patch level, which involves a dual consideration of data consistency and architecture consistency. To guarantee production efficiency, machines are operated steadily which can be modeled as stationary processes, while speech generally possesses more variations over time. It is believed that machine audio is more information-sparse than speech, which is similar to the difference between vision and language~\cite{he2022masked}. On one hand, fine-tuning audio classification models yields better consistency between pre-training data and fine-tuning data. On the other hand, modeling from patch level is more consistent with the inductive bias of machine audio than modeling from frame level. Here patch level refers to modeling the patches in audio spectrograms, while frame level refers to modeling the time frames in waveforms or spectrograms. Recent works have demonstrated that modeling the audio from patch level generally has superior performances in audio classification tasks to modeling it from frame level~\cite{li2024self,chen2024eat}. Thus, a pre-trained model with patch-level representation is adopted, and we will empirically demonstrate its superiority for the machine ASD task.

In this paper, we propose AnoPatch, in which a ViT~\cite{dosovitskiy2021an} backbone is trained to extract patch-level representations from mel-spectrograms. The parameters of the ViT backbone are initialized by a pre-trained audio model, and the backbone is further fine-tuned on machine audio by classifying the metadata associated with the machine (e.g. machine type, entity ID, speed). During detection, all patch-level representations are merged into a general embedding for each clip, and KNN~\cite{ramaswamy2000efficient} is applied to the embedding for anomaly detection.

Extensive experiments are conducted on the datasets of DCASE 2020 Task 2~\cite{Koizumi_DCASE2020_01} and DCASE 2023 Task 2~\cite{Dohi_arXiv2023_01}, both of which are machine ASD datasets with multiple machine types. AnoPatch achieves state-of-the-art (SOTA) performances on both datasets without leveraging model ensembling. We also compare multiple pre-trained speech and audio models and empirically demonstrate that consistency is crucial when adapting pre-trained models to the machine ASD task.

The main contributions of this paper can be summarized:

\begin{itemize}[leftmargin=2em]
  \item We propose AnoPatch which fine-tunes a pre-trained ViT backbone for the machine ASD task in consideration of better consistency.
  \item AnoPatch achieves SOTA performances on both the DCASE 2020 dataset and the DCASE 2023 dataset.
  \item We empirically demonstrate that pre-trained models with better consistency tend to possess better ASD capability.
\end{itemize}

\section{Related Work}

\subsection{Pre-trained Models for Speech and Audio Tasks}

A significant amount of pre-trained models have emerged in recent years, continuously setting up new SOTA performances on various speech and audio tasks. Most models have a focal point on either speech tasks or audio tasks, although some may generalize on both domains.

As for models focusing on speech tasks, Wav2Vec 2.0~\cite{baevski2020wav2vec} quantizes raw waveforms into discrete units and models them by a transformer-based encoder. HuBERT~\cite{hsu2021hubert} generates pseudo labels by clustering on mel spectrograms while adopting the network of Wav2Vec 2.0. UniSpeech~\cite{wang2021unispeech} combines contrastive learning with supervised learning. WavLM~\cite{chen2022wavlm} improves HuBERT by applying multiple data augmentations.

As for models focusing on audio tasks~\cite{gemmeke2017audio}, AST~\cite{gong21b_interspeech} fine-tunes a ViT model which is pre-trained on images for audio classification. BEATs~\cite{pmlr-v202-chen23ag} iteratively trains a ViT backbone and an acoustic tokenizer in which the acoustic tokenizer assigns pseudo labels for unlabeled data. Imagebind~\cite{girdhar2023imagebind} employs a ViT encoder to align audio with multiple modalities. Recently both ATST~\cite{li2024self} and EAT~\cite{chen2024eat} utilize self-teaching techniques to unsupervisedly train a ViT model.

It is noted that speech models tend to model speech by frames, while audio models tend to model audio by patches.

\subsection{Approaches for Machine ASD}

Approaches for machine ASD can be roughly divided into disentangled approaches and end-to-end approaches. Disentangled approaches~\cite{hou2023decoupling} employ neural networks to extract semantic features of machine audio and apply statistical anomaly detection algorithms to these features. Commonly used detection algorithms are KNN~\cite{ramaswamy2000efficient}, LOF~\cite{breunig2000lof}, and Gaussian mixture models (GMM). End-to-end models employ neural networks to directly estimate the anomaly score. Amongst all end-to-end models, autoencoder~\cite{Koizumi_DCASE2020_01,jiang2023unsupervised} assumes that anomalous samples may have bigger reconstruction error when trained to reconstruct normal samples, and flow model~\cite{dohi2021flow} directly models the normal distribution and estimates the likelihood.

\subsection{Pre-trained Models for Machine ASD}

Compared to the prevalent applications of pre-trained models for speech and audio tasks, few works explored the use of pre-trained models for machine ASD, probably because pre-trained features have not shown much superiority against pre-defined features. Previous works ~\cite{wilkinghoff2021sub,Daniluk2020} explored the use of OpenL3~\cite{cramer2019look} embeddings for the machine ASD task, yet none of which produced satisfactory results. Recently Han et al.~\cite{han2024exploring} explored the use of multiple pre-trained speech models~\cite{baevski2020wav2vec,hsu2021hubert,wang2021unispeech,chen2022wavlm}, which outperform multiple systems in the DCASE 2023 challenge.

We believe such a result may result from the difference between speech and machine audio. As pointed out in Section~\ref{sec:intro}, machine audio is more stationary and information-sparse compared to speech, which is more suitable for modeling it from patch level. On one hand, the parameters obtained from pre-training on speech data may not scale well on machine audio. On the other hand, since the inductive bias is different, the architecture suitable for speech tasks may not apply well to the machine ASD task. Therefore, we propose to fine-tune pre-trained models that are intended for audio classification and leverage patch-level representation.

\begingroup
\setlength{\tabcolsep}{3.7pt}
\renewcommand{\arraystretch}{1.3}
\begin{table*}[th]
  \caption{Comparison on the DCASE 2020 dataset between AnoPatch and previous SOTA models}
  \label{tab:dcase20_sota}
  \centering
  \scriptsize
  \begin{tabular}{ccccccc|c|cccccc|c|c}
    \toprule
    \multirow{2}*{Models} & \multicolumn{7}{c}{Development set} & \multicolumn{7}{c|}{Evaluation set} & All \\
    & fan & pump & slider & ToyCar & ToyConveyor & valve & mean & fan & pump & slider & ToyCar & ToyConveyor & valve & mean & mean \\
    \midrule
    No. 1~\cite{Giri2020} & 80.65 & 83.27 & 93.41 & 92.72 & 73.28 & 94.30 & 86.27 & 89.42 & 87.69 & 93.68 & 92.04 & 82.27 & 93.51 & 89.77 & 88.02 \\
    Sub-Cluster AdaCos~\cite{wilkinghoff2021sub} & 82.77 & 91.81 & 98.59 & 94.01 & \textbf{76.77} & 96.63 & 90.10 & 95.42 & 92.53 & 93.54 & 93.96 & \textbf{84.94} & \textbf{97.31} & 92.95 & 91.53 \\
    MFN~\cite{hou2023decoupling} & 83.71 & 90.82 & 98.70 & 91.97 & 71.29 & 96.49 & 88.83 & 94.72 & 92.94 & 97.58 & 94.31 & 77.54 & 94.88 & 92.00 & 90.38 \\
    STgram~\cite{liu2022anomalous} & \textbf{91.51} & 86.85 & 98.58 & 91.06 & 69.09 & \textbf{99.04} & 89.35 & - & - & - & - & - & - & - & - \\
    AnoPatch (Ours) & 86.46 & \textbf{93.10} & \textbf{99.20} & \textbf{96.10} & 73.20 & 97.53 & \textbf{90.93} & \textbf{95.56} & \textbf{94.34} & \textbf{99.77} & \textbf{96.00} & 83.74 & 96.26 & \textbf{94.28} & \textbf{92.58} \\
    \bottomrule
  \end{tabular}
\end{table*}
\endgroup

\begingroup
\setlength{\tabcolsep}{2pt}
\renewcommand{\arraystretch}{1.3}
\begin{table*}[th]
  \caption{Comparison on the DCASE 2023 dataset between AnoPatch and previous SOTA models}
  \label{tab:dcase23_sota}
  \centering
  \scriptsize
  \begin{tabular}{cccccccc|c|ccccccc|c|c}
    \toprule
    \multirow{2}*{Models} & \multicolumn{8}{c}{Development set} & \multicolumn{8}{c|}{Evaluation set} & All \\
    & bearing & fan & gearbox & slider & ToyCar & ToyTrain & valve & hmean & bandsaw & grinder & shaker & ToyDrone & ToyNscale & ToyTank & Vacuum & hmean & hmean \\
    \midrule
    No. 1~\cite{JieIESEFPT2023} & 64.41 & \textbf{76.27} & \textbf{74.78} & \textbf{91.83} & 51.66 & 53.17 & 85.44 & \textbf{68.11} & 60.97 & \textbf{65.18} & 63.50 & 55.71 & 84.92 & 60.72 & \textbf{92.27} & 66.97 & 67.54 \\
    No. 2~\cite{LvHUAKONG2023} & \textbf{72.39} & 62.41 & 74.41 & 87.84 & 59.10 & 58.67 & 65.53 & 67.38 & 55.47 & 64.76 & 70.98 & 52.89 & 71.90 & 70.73 & 91.48 & 66.39 & 66.88 \\
    Han et al.~\cite{han2024exploring} & 57.10 & 62.76 & 67.52 & 79.11 & \textbf{63.47} & 57.35 & \textbf{67.79} & 64.31 & - & - & - & - & - & - & - & - & -\\
    FeatEx~\cite{wilkinghoff2024self} & - & - & - & - & - & - & - & 66.95 & - & - & - & - & - & - & - & 68.52 & 67.73 \\
    AnoPatch (Ours) & 70.43 & 66.65 & 58.67 & 81.88 & 58.78 & \textbf{67.16} & 53.73 & 64.24 & \textbf{69.71} & 64.06 & \textbf{80.25} & \textbf{64.49} & \textbf{85.04} & \textbf{72.60} & 92.24 & \textbf{74.23} & \textbf{68.87} \\
    \bottomrule
  \end{tabular}
\end{table*}
\endgroup

\begingroup
\renewcommand{\arraystretch}{1.3}
\begin{table*}[th]
  \caption{Comparison on the DCASE 2020 dataset between multiple pre-trained models}
  \label{tab:dcase20_pretrain}
  \centering
  \begin{tabular}{cccc|cccccc|c}
    \toprule
    Set & Domain & Models & Size & fan & pump & slider & ToyCar & ToyConveyor & valve & mean \\
    \midrule
    \multirow{8}*{dev} & \multirow{4}*{Speech} & Wav2Vec 2.0~\cite{baevski2020wav2vec} & 316M & 79.04 & 80.16 & 89.55 & 94.04 & 70.11 & 92.49 & 84.23 \\
    & & HuBERT~\cite{hsu2021hubert} & 316M & 75.81 & 84.99 & 85.13 & 92.44 & 67.91 & 92.37 & 83.11 \\
    & & UniSpeech~\cite{wang2021unispeech} & 316M & 77.13 & 84.64 & 91.31 & 92.41 & 71.87 & 93.51 & 85.14 \\
    & & WavLM~\cite{chen2022wavlm} & 316M & 79.37 & 85.52 & 86.96 & 92.93 & 69.23 & 94.69 & 84.78 \\
    \cline{2-11}
    & \multirow{4}*{Audio} & AST~\cite{gong21b_interspeech} & 86M & 80.51 & 82.34 & 98.21 & 91.37 & 61.15 & 97.55 & 85.19 \\
    & & Imagebind~\cite{girdhar2023imagebind} & 86M & 82.44 & 87.98 & \textbf{98.41} & 92.24 & 68.86 & 91.42 & 86.89 \\
    & & ATST~\cite{li2024self} & 85M & 76.22 & 82.81 & 89.46 & 94.85 & \textbf{72.02} & 95.34 & 85.12 \\
    & & BEATs~\cite{pmlr-v202-chen23ag} (Adopted) & 90M & \textbf{84.01} & \textbf{92.87} & 94.72 & \textbf{96.55} & 71.42 & \textbf{99.62} & \textbf{89.87} \\
    \midrule
    \midrule
    \multirow{8}*{eval} & \multirow{4}*{Speech} & Wav2Vec 2.0~\cite{baevski2020wav2vec} & 316M & 85.67 & 81.03 & 87.86 & 93.23 & 82.29 & 93.76 & 87.31 \\
    & & HuBERT~\cite{hsu2021hubert} & 316M & 82.42 & 76.76 & 95.69 & 92.30 & 78.51 & 95.61 & 86.88\\
    & & UniSpeech~\cite{wang2021unispeech} & 316M & 83.33 & 79.39 & 92.98 & 93.96 & 81.91 & 91.89 & 87.24 \\
    & & WavLM~\cite{chen2022wavlm} & 316M & 84.35 & 77.72 & 94.53 & 92.40 & 77.13 & 90.16 & 86.05 \\
    \cline{2-11}
    & \multirow{4}*{Audio} & AST~\cite{gong21b_interspeech} & 86M & 90.68 & 86.08 & 93.92 & 90.65 & 63.91 & 92.14 & 86.23 \\
    & & Imagebind~\cite{girdhar2023imagebind} & 86M & 91.56 & 88.92 & 95.65 & 92.36 & 70.74 & 87.27 & 87.75 \\
    & & ATST~\cite{li2024self} & 85M & 86.76 & 84.11 & 88.07 & 94.04 & 79.87 & \textbf{96.87} & 88.29 \\
    & & BEATs~\cite{pmlr-v202-chen23ag} (Adopted) & 90M & \textbf{93.98} & \textbf{93.83} & \textbf{98.05} & \textbf{96.95} & \textbf{84.24} & 94.53 & \textbf{93.60} \\
    \bottomrule
  \end{tabular}
\end{table*}
\endgroup

\section{Proposed Method}

\begin{figure}
  \centering\includegraphics[width=0.8\linewidth]{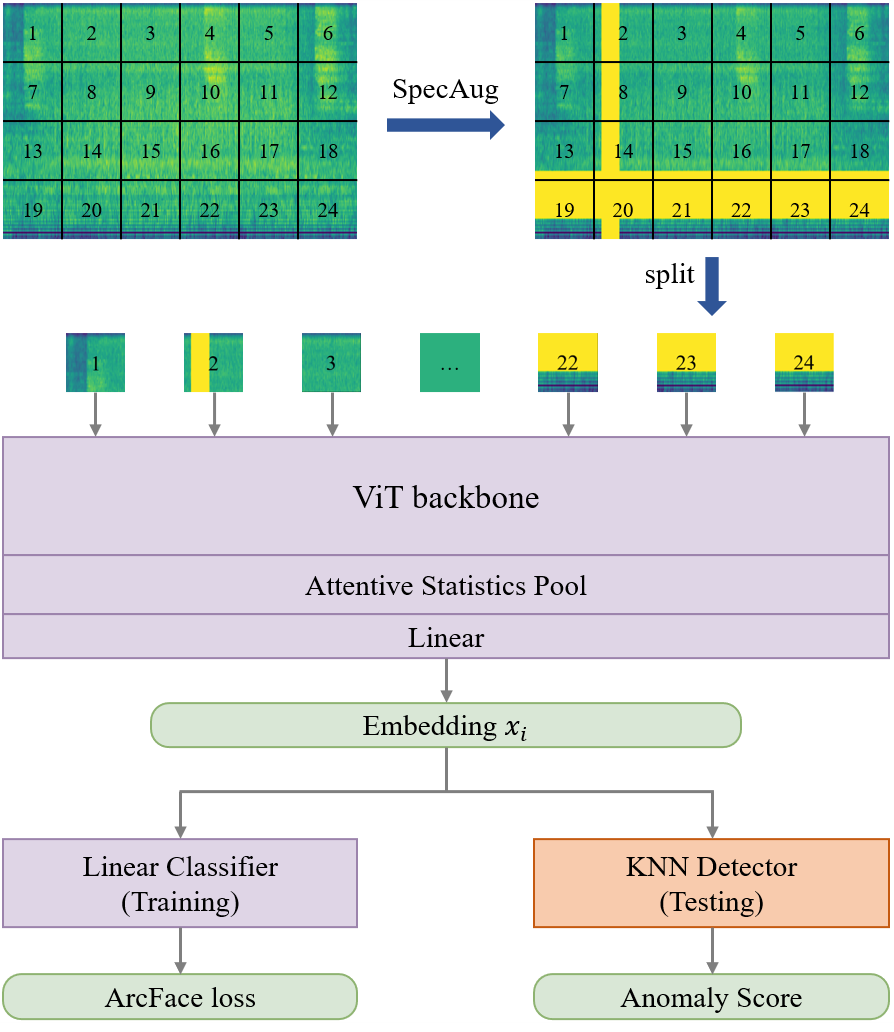}
  \caption{Architecture of AnoPatch}
  \label{fig:framework}
\end{figure}

\subsection{Backbone}

Figure~\ref{fig:framework} illustrates the overall architecture of AnoPatch, in which a ViT backbone is adopted to extract patch-level representations of machine audio. The parameters of the ViT backbone are initialized from BEATs~\cite{pmlr-v202-chen23ag} in consideration of better consistency between pre-training and fine-tuning, since BEATs is pre-trained on AudioSet for audio classification purposes. The input of AnoPatch is a mel-spectrogram which is then split into 16 $\times$ 16 patches. The ViT backbone models each patch as a single token and outputs an embedding for each patch. An attentive statistics pooling layer from ECAPA-TDNN~\cite{dawalatabad21_interspeech} is employed to merge patch embeddings into an utterance embedding $u_i$. Finally, the utterance embedding $u_i$ is linearly mapped to a low-dimensional embedding $x_i$ which is further utilized for anomaly detection.

\subsection{Fine-Tuning}

AnoPatch is fine-tuned on machine audio by classifying the metadata associated with the machine, in which each unique combination of the metadata is considered as a new label. Since the classification task is rather easy, ArcFace loss~\cite{deng2019arcface} is adopted to reinforce the task, which can be formulated as follows:

\begin{equation}
  L=-\frac{1}{N}\sum\limits_{i=1}^N\log\frac{e^{s\cos(\theta_{y_i}+m)}}{e^{s\cos(\theta_{y_i}+m)}+\sum\limits_{j=1,j{\neq}y_i}^{c}e^{s\cos\theta_{j}}}
\end{equation}
where $y_i$ is the label of sample $i$, and $s$ and $m$ are two hyperparameters. $\theta_j$ is the angle between the embedding of sample $i$ and the registered embedding of the j-th class, which is the j-th column of the weight $W$ of the classification head:

\begin{equation}
  \theta_j=\arccos\left(\frac{W_j^Tx_i}{{\lVert W_j \rVert}_2{\cdot}{\lVert x_i \rVert}_2}\right)
\end{equation}
where $x_i$ is the final output of the backbone, $T$ represents the transpose operation, and $\lVert{\cdot}\rVert$ denotes the L2 norm distance.

To further strengthen the fine-tuning task, SpecAug~\cite{park19e_interspeech} is adopted before splitting the spectrogram into patches, which is consistent with BEATs. It is noted that time warping is removed since it may be associated with anomalies.

\subsection{Anomaly Detection}

KNN~\cite{ramaswamy2000efficient} is adopted as the anomaly detection backend. Embeddings of normal samples form a memory bank, and the cosine distance between a query embedding and its closest neighbor is utilized as the anomaly score ($k=1$). For the DCASE 2020 dataset, KNN is individually applied to clips of the same machine type and machine ID. For the DCASE 2023 dataset, KNN is individually applied to each machine type, and we also adopt a soft scoring mechanism~\cite{Dohi_arXiv2023_01} where two KNN detectors are initialized by source samples and target samples respectively, and the minimum score given by two detectors is selected.

\section{Experiment}

\subsection{Datasets}

Experiments are conducted on the datasets of DCASE 2020 Task 2~\cite{Koizumi_DCASE2020_01} and DCASE 2023 Task 2~\cite{Dohi_arXiv2023_01}, which feature 7 and 14 machine types respectively. The DCASE 2020 dataset can be divided into a development set and an evaluation set, both of which contain a training subset with only normal audio clips, and a test subset with normal and anomalous audio clips. Machine type and machine ID are provided as metadata, where machine ID corresponds to different entities of a type of machine. The DCASE 2023 dataset shares a similar structure but introduces domain shift to reflect diverse working conditions. That is, for the training subset of each machine type, 990 clips are from the source domain, while only 10 clips are from the target domain. Machine types and attributes of working conditions are provided as metadata. 

Performances are evaluated by the area under the receiver operating characteristic (ROC) curve (AUC) and the partial-AUC (pAUC) following the challenge rules~\cite{Koizumi_DCASE2020_01,Dohi_arXiv2023_01}. For the DCASE 2020 dataset, we report the arithmetic mean of all AUC and pAUC for each machine type, and an arithmetic mean of the whole set. For the DCASE 2023 dataset, we report the harmonic mean of all AUC and pAUC for each machine type, and a harmonic mean of the whole set.

\subsection{Baseline Systems}

Multiple SOTA models are adopted as baselines. For the DCASE 2020 dataset, we adopt the best system of the challenge~\cite{Giri2020} and three single models~\cite{wilkinghoff2021sub,hou2023decoupling,liu2022anomalous} as baselines. Since MobileFaceNet (MFN)~\cite{chen2018mobilefacenets} is frequently adopted in previous literature~\cite{liu2022anomalous,hou2023decoupling} with competitive performances, we implement it on the DCASE 2020 dataset as a decent baseline. For the DCASE 2023 dataset, we adopt the top two systems of the challenge and two single models~\cite{han2024exploring,wilkinghoff2024self} as baselines. It is noted that most top-tier systems of both challenges are ensemble models, while AnoPatch is a single model.

\subsection{Implementation Details}

\begin{table}
  \caption{Detailed hyperparameter settings}
  \label{tab:setting}
  \centering
  \begin{tabular}{ccc}
    \toprule
    Items & DCASE 2020 & DCASE 2023 \\
    \midrule
    Number of classes & 41 & 167 \\
    SpecAug-freq & 40 & 80 \\
    SpecAug-time & 80 & 80 \\
    Total steps & \multicolumn{2}{c}{10k} \\
    Batch size & \multicolumn{2}{c}{32} \\
    Gradient accumulation & \multicolumn{2}{c}{8} \\
    Optimizer & \multicolumn{2}{c}{AdamW~\cite{loshchilov2018decoupled}} \\
    Learning rate & \multicolumn{2}{c}{1e-4} \\
    Warmup steps & \multicolumn{2}{c}{960} \\
    \bottomrule
  \end{tabular}
\end{table}

The ViT backbone consists of 12 layers with 90M parameters in total, which is initialized from BEATs (iteration 3 of pre-training). Identical with BEATs, the audio waveform is converted to a mel-spectrogram with a windows size of 25ms, a hop size of 10ms, and 128 Mel bins, and then the network transforms the whole spectrogram into a 128-dim embedding. Each unique combination of machine type and machine ID is regarded as a unique class for the DCASE 2020 dataset, while each unique combination of machine type and attributes is regarded as a unique class for the DCASE 2023 dataset. AnoPatch is trained by data of all machine types, and all parameters are learnable. Experiments are conducted separately on two datasets on an NVIDIA RTX A6000 GPU. Detailed hyperparameter settings are listed in Table~\ref{tab:setting}. 

\subsection{Experiment Results}

Table~\ref{tab:dcase20_sota} compares AnoPatch with previous SOTA models on the DCASE 2020 dataset. AnoPatch achieves the best performance with a mean of 92.58 on both sets, setting up a new milestone of 94.28 on the evaluation set.

Table~\ref{tab:dcase23_sota} compares AnoPatch with previous SOTA models on the DCASE 2023 dataset. As expected, AnoPatch outperforms all baseline models with a mean of 68.87 on both sets, also setting up a new milestone of 74.23 on the evaluation set.


\subsection{Comparison between Pre-trained Models}

Multiple large pre-trained models for speech and audio tasks are compared on the machine ASD task. Four speech models are employed, namely Wav2Vec 2.0~\cite{baevski2020wav2vec}, HuBERT~\cite{hsu2021hubert}, UniSpeech~\cite{wang2021unispeech}, and WavLM~\cite{chen2022wavlm}, all of which are pre-trained on speech datasets with a network suitable for grasping the dense semantics in speech, inheriting data inconsistency and architecture inconsistency. Four audio models are employed, namely AST~\cite{gong21b_interspeech}, Imagebind~\cite{girdhar2023imagebind}, ATST~\cite{li2024self}, and BEATs~\cite{pmlr-v202-chen23ag}, all of which are trained on audio datasets and model the audio from patch level. It is noted that AST is initialized from a ViT model pre-trained on images, thus inheriting data inconsistency. All models are fine-tuned by the same labels and ArcFace loss. Four speech models are implemented following~\cite{han2024exploring}, while weighted sum, status augmentation, and transformer pooling are removed for fair comparison. The fine-tuning processes of three other audio models are mainly the same as BEATs. If the audio model has incorporated a special token that aggregates the features over time (e.g. [cls] token), this token is adopted as the utterance embedding $u_i$. If not, an attentive statistics pooling layer is appended to the model to aggregate the features. It is noted that SpecAug is also removed for a fair comparison.

Table~\ref{tab:dcase20_pretrain} compares the performances of these pre-trained models on the DCASE 2020 dataset. As for the development set, the best performance of speech models is 85.14, while the worst performance of audio models is 85.19. As for the evaluation set, all audio models except AST outperform the best speech models, which can be explained by the data inconsistency of AST. The improvement even goes up to 4.73 and 6.29 on two subsets. Thus, it is believed that pre-trained audio models leveraging patch-level representation shall have superior performances for the machine ASD task.

\section{Conclusion}

In this paper, we focused on adapting large pre-trained models to the machine ASD task. It was believed that the inconsistency between pre-training and fine-tuning deprecates the performances of previous attempts, which was empirically proved by the comparison in Table~\ref{tab:dcase20_pretrain}. Therefore, we proposed AnoPatch which contains a ViT backbone initialized by BEATs. As a result, AnoPatch demonstrates SOTA results on the DCASE 2020 dataset and the DCASE 2023 dataset.

\section{Acknowledgements}

This work was supported by the National Key Research and Development Program of China (Grant NO.2021YFA1000500(4)) and the National Natural Science Foundation of China under Grant No. 62276153.

\bibliographystyle{IEEEtran}
\bibliography{mybib}

\begin{thebibliography}{10}
\providecommand{\url}[1]{#1}
\csname url@samestyle\endcsname
\providecommand{\newblock}{\relax}
\providecommand{\bibinfo}[2]{#2}
\providecommand{\BIBentrySTDinterwordspacing}{\spaceskip=0pt\relax}
\providecommand{\BIBentryALTinterwordstretchfactor}{4}
\providecommand{\BIBentryALTinterwordspacing}{\spaceskip=\fontdimen2\font plus
\BIBentryALTinterwordstretchfactor\fontdimen3\font minus \fontdimen4\font\relax}
\providecommand{\BIBforeignlanguage}[2]{{%
\expandafter\ifx\csname l@#1\endcsname\relax
\typeout{** WARNING: IEEEtran.bst: No hyphenation pattern has been}%
\typeout{** loaded for the language `#1'. Using the pattern for}%
\typeout{** the default language instead.}%
\else
\language=\csname l@#1\endcsname
\fi
#2}}
\providecommand{\BIBdecl}{\relax}
\BIBdecl

\bibitem{wilkinghoff2021sub}
K.~Wilkinghoff, ``Sub-cluster adacos: Learning representations for anomalous sound detection,'' in \emph{2021 International Joint Conference on Neural Networks (IJCNN)}.\hskip 1em plus 0.5em minus 0.4em\relax IEEE, 2021, pp. 1--8.

\bibitem{liu2022anomalous}
Y.~Liu, J.~Guan, Q.~Zhu, and W.~Wang, ``Anomalous sound detection using spectral-temporal information fusion,'' in \emph{ICASSP 2022-2022 IEEE International Conference on Acoustics, Speech and Signal Processing (ICASSP)}.\hskip 1em plus 0.5em minus 0.4em\relax IEEE, 2022, pp. 816--820.

\bibitem{hou2023decoupling}
Q.~Hou, A.~Jiang, W.-Q. Zhang, P.~Fan, and J.~Liu, ``Decoupling detectors for scalable anomaly detection in aiot systems with multiple machines,'' in \emph{2023 IEEE Global Communications Conference (GLOBECOM)}.\hskip 1em plus 0.5em minus 0.4em\relax IEEE, 2023, pp. 1--6.

\bibitem{he2022masked}
K.~He, X.~Chen, S.~Xie, Y.~Li, P.~Doll{\'a}r, and R.~Girshick, ``Masked autoencoders are scalable vision learners,'' in \emph{Proceedings of the IEEE/CVF conference on computer vision and pattern recognition}, 2022, pp. 16\,000--16\,009.

\bibitem{li2024self}
X.~Li, N.~Shao, and X.~Li, ``Self-supervised audio teacher-student transformer for both clip-level and frame-level tasks,'' \emph{IEEE/ACM Transactions on Audio, Speech, and Language Processing}, 2024.

\bibitem{chen2024eat}
W.~Chen, Y.~Liang, Z.~Ma, Z.~Zheng, and X.~Chen, ``Eat: Self-supervised pre-training with efficient audio transformer,'' in \emph{Proceedings of the 33rd International Joint Conference on Artificial Intelligence}, 2024.

\bibitem{dosovitskiy2021an}
A.~Dosovitskiy, L.~Beyer, A.~Kolesnikov, D.~Weissenborn, X.~Zhai, T.~Unterthiner, M.~Dehghani, M.~Minderer, G.~Heigold, S.~Gelly, J.~Uszkoreit, and N.~Houlsby, ``An image is worth 16x16 words: Transformers for image recognition at scale,'' in \emph{International Conference on Learning Representations}, 2021.

\bibitem{ramaswamy2000efficient}
S.~Ramaswamy, R.~Rastogi, and K.~Shim, ``Efficient algorithms for mining outliers from large data sets,'' in \emph{Proc. 2000 ACM SIGMOD Int. Conf. Manag. Data}, 2000, pp. 427--438.

\bibitem{Koizumi_DCASE2020_01}
Y.~Koizumi, Y.~Kawaguchi, K.~Imoto, T.~Nakamura, Y.~Nikaido, R.~Tanabe, H.~Purohit, K.~Suefusa, T.~Endo, M.~Yasuda, and N.~Harada, ``Description and discussion on {DCASE}2020 challenge task2: Unsupervised anomalous sound detection for machine condition monitoring,'' in \emph{Proceedings of the Detection and Classification of Acoustic Scenes and Events 2020 Workshop (DCASE2020)}, November 2020, pp. 81--85.

\bibitem{Dohi_arXiv2023_01}
K.~Dohi, K.~Imoto, N.~Harada, D.~Niizumi, Y.~Koizumi, T.~Nishida, H.~Purohit, R.~Tanabe, T.~Endo, and Y.~Kawaguchi, ``Description and discussion on {DCASE} 2023 challenge task 2: First-shot unsupervised anomalous sound detection for machine condition monitoring,'' \emph{In arXiv e-prints: 2305.07828}, 2023.

\bibitem{baevski2020wav2vec}
A.~Baevski, Y.~Zhou, A.~Mohamed, and M.~Auli, ``wav2vec 2.0: A framework for self-supervised learning of speech representations,'' \emph{Advances in neural information processing systems}, vol.~33, pp. 12\,449--12\,460, 2020.

\bibitem{hsu2021hubert}
W.-N. Hsu, B.~Bolte, Y.-H.~H. Tsai, K.~Lakhotia, R.~Salakhutdinov, and A.~Mohamed, ``Hubert: Self-supervised speech representation learning by masked prediction of hidden units,'' \emph{IEEE/ACM Transactions on Audio, Speech, and Language Processing}, vol.~29, pp. 3451--3460, 2021.

\bibitem{wang2021unispeech}
C.~Wang, Y.~Wu, Y.~Qian, K.~Kumatani, S.~Liu, F.~Wei, M.~Zeng, and X.~Huang, ``Unispeech: Unified speech representation learning with labeled and unlabeled data,'' in \emph{International Conference on Machine Learning}.\hskip 1em plus 0.5em minus 0.4em\relax PMLR, 2021, pp. 10\,937--10\,947.

\bibitem{chen2022wavlm}
S.~Chen, C.~Wang, Z.~Chen, Y.~Wu, S.~Liu, Z.~Chen, J.~Li, N.~Kanda, T.~Yoshioka, X.~Xiao \emph{et~al.}, ``Wavlm: Large-scale self-supervised pre-training for full stack speech processing,'' \emph{IEEE Journal of Selected Topics in Signal Processing}, vol.~16, no.~6, pp. 1505--1518, 2022.

\bibitem{gemmeke2017audio}
J.~F. Gemmeke, D.~P. Ellis, D.~Freedman, A.~Jansen, W.~Lawrence, R.~C. Moore, M.~Plakal, and M.~Ritter, ``Audio set: An ontology and human-labeled dataset for audio events,'' in \emph{2017 IEEE international conference on acoustics, speech and signal processing (ICASSP)}.\hskip 1em plus 0.5em minus 0.4em\relax IEEE, 2017, pp. 776--780.

\bibitem{gong21b_interspeech}
Y.~Gong, Y.-A. Chung, and J.~Glass, ``{AST: Audio Spectrogram Transformer},'' in \emph{Proc. Interspeech 2021}, 2021, pp. 571--575.

\bibitem{pmlr-v202-chen23ag}
S.~Chen, Y.~Wu, C.~Wang, S.~Liu, D.~Tompkins, Z.~Chen, W.~Che, X.~Yu, and F.~Wei, ``{BEAT}s: Audio pre-training with acoustic tokenizers,'' in \emph{Proceedings of the 40th International Conference on Machine Learning}, ser. Proceedings of Machine Learning Research, vol. 202.\hskip 1em plus 0.5em minus 0.4em\relax PMLR, 23--29 Jul 2023, pp. 5178--5193.

\bibitem{girdhar2023imagebind}
R.~Girdhar, A.~El-Nouby, Z.~Liu, M.~Singh, K.~V. Alwala, A.~Joulin, and I.~Misra, ``Imagebind: One embedding space to bind them all,'' in \emph{Proceedings of the IEEE/CVF Conference on Computer Vision and Pattern Recognition}, 2023, pp. 15\,180--15\,190.

\bibitem{breunig2000lof}
M.~M. Breunig, H.-P. Kriegel, R.~T. Ng, and J.~Sander, ``Lof: identifying density-based local outliers,'' in \emph{Proc. 2000 ACM SIGMOD Int. Conf. Manag. Data}, 2000, pp. 93--104.

\bibitem{jiang2023unsupervised}
A.~Jiang, W.-Q. Zhang, Y.~Deng, P.~Fan, and J.~Liu, ``Unsupervised anomaly detection and localization of machine audio: A gan-based approach,'' in \emph{ICASSP 2023-2023 IEEE International Conference on Acoustics, Speech and Signal Processing (ICASSP)}.\hskip 1em plus 0.5em minus 0.4em\relax IEEE, 2023, pp. 1--5.

\bibitem{dohi2021flow}
K.~Dohi, T.~Endo, H.~Purohit, R.~Tanabe, and Y.~Kawaguchi, ``Flow-based self-supervised density estimation for anomalous sound detection,'' in \emph{ICASSP 2021-2021 Proc. IEEE Int. Conf. Acoust., Speech, Signal Process. (ICASSP)}.\hskip 1em plus 0.5em minus 0.4em\relax IEEE, 2021, pp. 336--340.

\bibitem{Daniluk2020}
P.~Daniluk, M.~Gozdziewski, S.~Kapka, and M.~Kosmider, ``Ensemble of auto-encoder based systems for anomaly detection,'' DCASE2020 Challenge, Tech. Rep., July 2020.

\bibitem{cramer2019look}
A.~L. Cramer, H.-H. Wu, J.~Salamon, and J.~P. Bello, ``Look, listen, and learn more: Design choices for deep audio embeddings,'' in \emph{ICASSP 2019-2019 IEEE International Conference on Acoustics, Speech and Signal Processing (ICASSP)}.\hskip 1em plus 0.5em minus 0.4em\relax IEEE, 2019, pp. 3852--3856.

\bibitem{han2024exploring}
B.~Han, Z.~Lv, A.~Jiang, W.~Huang, Z.~Chen, Y.~Deng, J.~Ding, C.~Lu, W.-Q. Zhang, P.~Fan, J.~Liu, and Y.~Qian, ``Exploring large scale pre-trained models for robust machine anomalous sound detection,'' in \emph{ICASSP 2024-2024 IEEE International Conference on Acoustics, Speech and Signal Processing (ICASSP)}.\hskip 1em plus 0.5em minus 0.4em\relax IEEE, 2024, pp. 1--5.

\bibitem{Giri2020}
R.~Giri, S.~V. Tenneti, K.~Helwani, F.~Cheng, U.~Isik, and A.~Krishnaswamy, ``Unsupervised anomalous sound detection using self-supervised classification and group masked autoencoder for density estimation,'' DCASE2020 Challenge, Tech. Rep., July 2020.

\bibitem{JieIESEFPT2023}
J.~Jie, ``Anomalous sound detection based on self-supervised learning,'' DCASE2023 Challenge, Tech. Rep., June 2023.

\bibitem{LvHUAKONG2023}
Z.~Lv, B.~Han, Z.~Chen, Y.~Qian, J.~Ding, and J.~Liu, ``Unsupervised anomalous detection based on unsupervised pretrained models,'' DCASE2023 Challenge, Tech. Rep., June 2023.

\bibitem{wilkinghoff2024self}
K.~Wilkinghoff, ``Self-supervised learning for anomalous sound detection,'' in \emph{ICASSP 2024-2024 IEEE International Conference on Acoustics, Speech and Signal Processing (ICASSP)}.\hskip 1em plus 0.5em minus 0.4em\relax IEEE, 2024, pp. 276--280.

\bibitem{dawalatabad21_interspeech}
N.~Dawalatabad, M.~Ravanelli, F.~Grondin, J.~Thienpondt, B.~Desplanques, and H.~Na, ``{ECAPA-TDNN Embeddings for Speaker Diarization},'' in \emph{Proc. Interspeech 2021}, 2021, pp. 3560--3564.

\bibitem{deng2019arcface}
J.~Deng, J.~Guo, N.~Xue, and S.~Zafeiriou, ``Arcface: Additive angular margin loss for deep face recognition,'' in \emph{Proceedings of the IEEE/CVF conference on computer vision and pattern recognition}, 2019, pp. 4690--4699.

\bibitem{park19e_interspeech}
D.~S. Park, W.~Chan, Y.~Zhang, C.-C. Chiu, B.~Zoph, E.~D. Cubuk, and Q.~V. Le, ``{SpecAugment: A Simple Data Augmentation Method for Automatic Speech Recognition},'' in \emph{Proc. Interspeech 2019}, 2019, pp. 2613--2617.

\bibitem{chen2018mobilefacenets}
S.~Chen, Y.~Liu, X.~Gao, and Z.~Han, ``Mobilefacenets: Efficient cnns for accurate real-time face verification on mobile devices,'' in \emph{Biometric Recognition: 13th Chinese Conference, CCBR 2018, Urumqi, China, August 11-12, 2018, Proceedings 13}.\hskip 1em plus 0.5em minus 0.4em\relax Springer, 2018, pp. 428--438.

\bibitem{loshchilov2018decoupled}
I.~Loshchilov and F.~Hutter, ``Decoupled weight decay regularization,'' in \emph{International Conference on Learning Representations}, 2019.

\end{thebibliography}

\end{document}